# Van der Waals Heterostructure Magnetic Josephson Junction

H. Idzuchi[1#], F. Pientka[1,2], K.-F. Huang[1], K. Harada[3], Ö. Gül[1], Y. J. Shin[1##], L. T. Nguyen[4], N. H. Jo[5,6], D. Shindo[3], R. J. Cava[4], P. C. Canfield[5,6], and P. Kim[1*]

[1] *Department of Physics, Harvard University, Cambridge, MA 02138, USA*

[2] *Institut für Theoretische Physik, Goethe-Universität, 60438 Frankfurt am Main, Germany*

[3] *Center for Emergent Matter Science (CEMS), RIKEN, Wako, Saitama 351-0198, Japan*

[4] *Department of Chemistry, Princeton University, Princeton, NJ 08540, USA*

[5] *Department of Physics and Astronomy, Iowa State University, Ames, IA 50011, USA*

[6] *Ames Laboratory, Iowa State University, Ames, IA 50011, USA*

[*]pkim@physics.harvard.edu

[#]current affiliation: Tohoku University

[##]current affiliation: Center for Functional Nanomaterials, Brookhaven National Laboratory, Upton, New York 11973, United States of America

**When two superconductors are connected across a ferromagnet, the spin configuration of the transferred Cooper pairs can be modulated due to magnetic exchange interaction. The resulting supercurrent can reverse its sign across the Josephson junction (JJ) [1-4]. Here we demonstrate Josephson phase modulation in van der Waals heterostructures when Cooper pairs from superconducting $NbSe_2$ tunnel through atomically thin magnetic insulator (MI) $Cr_2Ge_2Te_6$. Employing a superconducting quantum interference device based on MI JJs, we probe a doubly degenerate non-trivial JJ phase (ϕ) originating from the magnetic barrier. This ϕ–phase JJ is formed by momentum conserving tunneling of Ising Cooper pairs [5] across magnetic domains in the $Cr_2Ge_2Te_6$ barrier. The doubly degenerate ground states in MI JJs provide a two-level quantum system that can be utilized as a new disipationless component for superconducting quantum devices, including phase batteries [6], memories [7,8], and quantum Ratchets [9, 10].**

As a phase difference φ develops between two superconductors, a DC Josephson supercurrent $I_s = I_c \sin(φ)$ flows through the junction. The vanishing supercurrent at the minimum energy imposes the condition that only φ = 0 or π. For conventional superconductors with spin-singlet pairing, the spatially symmetric Cooper pair wavefunction enforces φ = 0 as the ground state. However, when the superconducting (S) electrodes are separated by a ferromagnetic barrier (F), Cooper pairs can acquire an additional phase when tunneling through the magnetic barrier, yielding a spatial oscillation of the superconducting order parameter in the barrier [2,11,12]. Carefully tuning the thickness of the F-layer, $d_F$, can reverse the sign of the superconducting order parameter across the barrier owing to an exchange-energy driven phase shift [1-4], resulting in a π-phase JJ. Engineering the system to achieve an arbitrary ϕ-phase between 0 and π is possible by coherently combining 0- and π- JJs [13]. For metallic F-JJ, for example, a long channel JJ (typically 100 μm), combining both 0 and π junctions exhibit ϕ–phase characteristics [14]. ϕ-phase JJs (ϕ-JJs) can serve as useful components for various superconducting quantum electronic devices, such as phase batteries that can be used to bias both classical and quantum circuits, superconducting-magnet hybrid memories and JJ-based quantum ratchets [6-10,15,16]. Often, a metallic ferromagnetic barrier requires $d_F \geq 5$ nm and a macroscopic lateral junction size, and is not in general suitable for use in dissipationless, compact quantum device components. JJs using magnetic semiconducting GdN barriers in the spin filter device geometry was constructed [17], exhibiting an unconventional second harmonic current-phase relation and switching characteristics [18,19]. However, the demonstration of Josephson phase engineering in dissipationless magnetic JJs yet to be realized.

Van der Waals (vdW) heterostructures are ideal platforms for creating atomically thin Josephson coupling systems. Here, we use a few atomic layers of the vdW magnetic insulator (MI) $Cr_2Ge_2Te_6$ [20] as the magnetic barrier to construct ϕ-JJs. Previous studies show that Josephson coupling can occur between cleaved surfaces of the vdW superconductor $NbSe_2$ [21]. Here we assembled $NbSe_2/Cr_2Ge_2Te_6/NbSe_2$ heterostructures (Fig. 1a), with a modified dry-transfer technique (see Methods for device fabrication). Our heterostructures clearly displays Josephson coupling for monolayer (ML) through 6 ML $Cr_2Ge_2Te_6$ barriers. Fig.1b-d shows the current density ($J$) versus voltage ($V$) characteristic across the JJs with 1-, 2-, and 6-ML MI barriers. For all devices, we find a clear Josephson supercurrent regime at low bias current, which turns into normal conduction at high bias current. The $J$-$V$ characteristic is hysteric indicating a switching current density $J_c$ (transition from superconducting to normal state) larger than the retrapping current density $J_r$ (transition from normal to superconducting state). $J_c$ becomes considerably larger than $J_r$ for thinner junctions, which is expected since the junction capacitance is larger for smaller $d_F$. For $J>J_c$, we obtain the normal state resistance $R_N = (1/A)dV/dJ$, where $A$ is the effective area of the junction. Fig.1e shows the comparison of $R_NA$ obtained from devices with three different $d_F$. We find an exponential increase of $R_NA$, fitted well to $a\exp(d_F/t)$, where the characteristic quasiparticle tunneling length is $t \approx 1.3$ nm and the normalized barrier resistance is $a \approx 340$ $\Omega$ $\mu m^2$. Importantly, our junction resistance is much lower than is displayed by one of a typical non-vdW ferromagnetic barrier such as EuS ($10^7 - 10^9$ $\Omega$ $\mu m^2$ for the thickness of 2.5 nm) [22]. This relatively small junction resistance is consistent with the smaller semiconducting energy gap of $Cr_2Ge_2Te_6$ (~0.4 eV in plane and ~ 1 eV out-of-plane) [23]. Interestingly, we find that while $R_NA$ increases exponentially with increasing $d_F$, the critical current density $J_c$ decreases more rapidly. Fig. 1e shows that the product $V_C=J_cR_NA$ decreases exponentially with increasing $d_F$, following $V_C = V_0\exp(-d_F/\xi_F)$ with the prefactor $V_0 \approx 0.8$ mV and a characteristic barrier tunneling length in $Cr_2Ge_2Te_6$ $\xi_F \approx 1.4$ nm. While $V_0$ is comparable to $V_C \sim 0.65$ mV in $NbSe_2$/Graphene/$NbSe_2$ junctions [24], the rapid decrease of $V_C$ with increasing $d_F$ indicates that the JJ coupling becomes weaker with a thicker magnetic barrier, as expected.

To demonstrate the role played by the ferromagnetism in $Cr_2Ge_2Te_6$, we measure the JJ critical current as a function of applied magnetic field. Figs. 2a-b show the in-plane and out-of-plane magnetic-field-dependent switching current $I_C = J_cA$ of the $NbSe_2/Cr_2Ge_2Te_6$(6ML)$/NbSe_2$ JJ. We observe hysteretic behavior of $I_C(H)$ for both field directions. $I_C(H)$ also shows a sudden drop near zero magnetic field. Interestingly, we find that the hysteresis in a magnetic field reaches values of ~ ±1.5 T, much larger than the field required to reach the saturation magnetization of our $Cr_2Ge_2Te_6$ bulk crystals (Fig. 2c). Furthermore, the magnetization of our bulk $Cr_2Ge_2Te_6$ shows neither a notable hysteresis nor a strong magnetic anisotropy, consistent with earlier reports [25]. The larger hysteresis field compared to the saturation magnetic field of $Cr_2Ge_2Te_6$ and the strong anisotropy observed in $I_C(H)$ thus cannot be simply attributed to the magnetization of $Cr_2Ge_2Te_6$ alone - rather, the large hysteresis loop for $I_C(H)$ can be related to the microscopic magnetic domain structure of $Cr_2Ge_2Te_6$. Using Lorentz transmission electron microscopy (see Methods for details), we find that thin $Cr_2Ge_2Te_6$ flakes develop two different magnetic domain structures: stripe-like (Fig. 2d) and bubble-like (Fig. 2e). The characteristic domain size is ~100 nm, consistent with previously reported multiple domain structures in thicker $Cr_2Ge_2Te_6$ flakes, where the stripe-phase is more stable than the metastable bubble phase [26]. We conclude that the interplay between the magnetic domains of $Cr_2Ge_2Te_6$ and the field-dependent Abrikosov vortex lattice in $NbSe_2$ can induce a transition between magnetic states and explain the experimental observations, including the sudden drop in $I_c(H)$ near zero magnetic field (Supplement Information).

The MI layer in JJs can create a nontrivial phase shift of the tunneling Cooper pairs. To probe this Josephson phase, we have realized SQUIDs consisting of one MI JJ ($NbSe_2/Cr_2Ge_2Te_6/NbSe_2$) and one reference JJ ($NbSe_2/NbSe_2$). After the assembly, we create the device by etching away the unnecessary areas (Fig.3a, note the edges of the $NbSe_2$ flakes were aligned parallel to each other (see Methods for details)). The wider MI JJ allows us to balance the critical currents for each JJ for a maximal SQUID critical current $I_{SQUID}^c(\Phi)$ as a function of magnetic flux $\Phi$ threaded through the SQUID loop. $I_{SQUID}^c(\Phi)$ measured in the SQUID (fig.3b) exhibits oscillations with the periodicity $\Phi_0 = h/2e$. However, we observe an irregular SQUID response in the field range between -1.2 mT and -2.2 mT, with a telegraph-like signal oscillating between two metastable critical current branches (Fig.3b) possibly related to the sudden change in critical current seen in Fig.2a. This bistable switching state is an indirect indication of a doubly degenerate ground state of the system. Nevertheless, the regular oscillation around zero magnetic field allows us to extract the phase of the MI JJ ($NbSe_2/Cr_2Ge_2Te_6/NbSe_2$). In a previous study of SQUIDs with ferromagnetic metallic spin valves [8], a controllable switching between 0- and π- Josephson junctions has been demonstrated. A SQUID that combines 0/0 or π/π JJs shows a maximal $I_{SQUID}^c(0)$ (defined as a 0-phase JJ), whereas a SQUID combining 0/π JJs shows a minimal $I_{SQUID}^c(0)$ (defined as a π- phase JJ).

For our SQUID with the MI JJ, we use two schemes to measure the two different switching currents (Fig. 3c): a switching current $I_{c-}$ obtained by sweeping from large negative bias to positive bias and another one, $I_{c0}$, obtained by sweeping from large positive bias to zero and then back to positive bias. Generally, we find $I_{c-} > I_{c0}$. More importantly, the phases of their oscillations are different, as shown in Fig. 3d. To obtain the absolute phase of $I_{SQUID}^c(\Phi)$, we have carefully calibrated the setting of the electromagnet that gives zero magnetic field, by using several on-chip Al SQUIDs with different sizes (see SI Fig. S3 for details). Strikingly, we find that none of the switching schemes provide 0 or π phase but $\phi_{c-}$= 259° and $\phi_{c0}$ = 59° as shown in Fig. 3d.

The presence of these two nontrivial phases (i.e. not simple multiples of 180 degrees) is reminiscent of two switching current states in a metallic ferromagnetic (F) $\phi$-JJ [13,14]. In those JJs, doubly degenerate ground states are realized by laterally connecting 0- and π- FJJs. Here, the π-JJ requires the thickness of the F-layer to be comparable to the wavelength of the order-parameter oscillation, implying $d_F$~ 10 nm, set by the exchange energy [12]. In addition, the formation of a $\phi$-FJJ in the 0-π JJ arrays needs a Josephson vortex pinned at the 0-π junction. This condition restricts the widths of the 0-JJ and π-JJ perpendicular to the supercurrent flow direction to be much longer than the Josephson length, $\lambda_J = \sqrt{\Phi_0/2\pi\mu_0 J_c \lambda_c}$ , where $\lambda_c$ is the magnetic penetration length. Such values of $d_F$ and $\lambda_J$ are incompatible with our atomically thin $Cr_2Ge_2Te_6$ based JJ. Specifically, our $Cr_2Ge_2Te_6$ barrier is an atomically thin insulator ($d_F$ = 1 nm), a case that is too thin to exhibit spatial order-parameter oscillations. Furthermore, the lateral size of our JJ, $L < 5$ μm, is much smaller than $\lambda_J$ ~10 μm, estimated using our experimentally obtained $J_c$ and $\lambda_c \approx 0.1$ μm for NbSe$_2$ reported previously [27]. Therefore, the observed $\phi$-JJ formation in our atomically thin MI-JJ, signaled by the appearance of doubly degenerate nontrivial phase shifts, requires an alternative mechanism for explanation.

The single-crystallinity of our vdW heterostructure combined with the strong spin orbit coupling (SOC) in the quasi-2D superconductor (S) constituent provides two new characteristics for Josephson coupling that are absent in conventional metallic F-JJs. First, in contrast to the F-JJs constructed by sputtered heterostructures [17], momentum-conserving tunneling in crystalline vdW heterostructures is allowed between the closely-aligned Fermi surfaces of two S-layers in vdW JJ, as the top and bottom S layers in our JJ are aligned along the same crystallographic axis (< 2°~5° misalignment; see Methods). Second, the strong SOC in NbSe$_2$ fixes the spin quantization axis of the Cooper pairs [5,28] normal to the substrate, denoted ↑ and ↓. In NbSe$_2$, due to weak interlayer tunneling and strong inversion symmetry breaking within each layer, two spin components remain localized predominantly in even or odd layers with a sizable spin splitting $\Delta_{SOC} \simeq 100$ meV [29]. This spin-layer locking results in unconventional Ising Cooper pairing *(K↑-K´↓* or *K↓-K´↑)* inside each layer where *K* and *K´* denote electronic band near *K* and *K´* points.

The Josephson phase between Ising Cooper pairs (ICPs) on the surfaces of NbSe$_2$ across the MI barrier can be sensitively modified by the magnetization direction. For out-of-plane magnetization, the spin of the ICPs is aligned parallel or anti-parallel with the spin of the MI. Similar to a previous theoretical study of JJs with magnetic impurities [30], the wave function of ICPs tunneling across the ferromagnetic junction can acquire an additional minus sign with respect to the nonmagnetic junction for sufficiently strong magnetic scattering (Fig. 4a, see Methods for details). Importantly, this sign flip of the Josephson coupled ICPs sets the phase of the JJ ground state to be $\varphi=\pi$. As the magnetization of the MI-layer is tilted away from the tunneling direction, the spin of the ICPs can flip during the tunneling process (see Methods and Supplementary Information for details). As a result, the ground state of the Josephson junction is at $\varphi=0$. (Fig. 4b). Unlike metallic F-JJs, our heterostructure allows for both 0- and π–JJs by adjusting the direction of the magnetization in the MI.

A parallel arrangement of 0- and π–JJ can lead to a degenerate $\phi$-JJ. Our MI-JJ offers such lateral arrays, created by magnetic domain structures in the MI, similar to what is shown in Fig. 2d-e. Here, the domains with out-of-plane magnetization separated by boundaries with titled spins could lead to a coexistence of 0 and π junction segments. Using a simple model based on a short junction with finite transparency *D* and a fraction of the plane λ that favors a π junction, we can estimate the Josephson energy $E_J$ of the junction, which provides two $\phi$ values for the degenerate ground states (see method section for more details). As an example, Fig. 4d shows $E_J(\phi)$ computed using this model with λ=0.53 and $D = 0.75$, resulting in the appearance of two ground states at different nontrivial Josephson phases $\phi_1 \simeq 100°$ and $\phi_2 \simeq 260°$.

Experimentally, the presence of two minimal phase angles in $E_J(\phi)$ can be directly revealed by measuring the JJ switching current distributions. This distribution is sensitively determined by the escape rate $\tau^{-1}$ from a tilted washboard potential that is created by biasing $E_J(\phi)$ with current $I$ (insets of Fig. 4g). Figs. 4e and 4f show the switching current distribution measured using the two above-mentioned sweep schemes, as a function of current $I(t)$ increasing monotonically with time $t$. The switching current distribution shows not only different values of the critical current but also a much wider distribution for the $\phi_{c0}$ state than for the $\phi_{c-}$ state (Fig.4f). Fig. 4g shows the escape rate for both ground states, which is calculated using the normalized distribution function $P(I_c)$ and the Fulton and Dunkleberger formula $\tau = (1 - \int_0^I P(u)du)/[P(I)\frac{dI}{dt}]$ [31]. Generally, we find the escape rate for the $\phi_{c0}$ state to be larger than for the $\phi_{c-}$ state, suggesting the $\phi_{c-}$ state is more stable than the $\phi_{c0}$ state under a bias current. Assuming thermal activation dominated escape gives $\tau^{-1} \sim e^{-\Delta/kT}$, where $\Delta$ is the barrier height in the tilted washboard potential and $k$ is the Boltzmann constant. Since the experimentally estimated $\tau^{-1}$ is smaller for $\phi_{c-}$ than that for $\phi_{c0}$, we infer that $\Delta_1 < \Delta_2$ (Fig. 4d), where $\Delta_1$ is responsible for the switching of $\phi_{c0}$ and $\Delta_2$ for the switching of $\phi_{c-}$, in agreement with the model presented in Fig. 4d. For an applied bias current $I$ smaller than the switching current of the $\phi_{c-}$ state, re-trapping is allowed (inset of Fig. 4g), which consistently explains the slower increase of $\tau^{-1}$ of the $\phi_{c0}$ state before the switching of the $\phi_{c-}$ state at higher current.

In conclusion, we have demonstrated the engineering of doubly degenerate $\phi$-JJs in a vdW system using an atomically thin NbSe$_2$ Ising superconductor and the domain structure in the ferromagnetic insulator Cr$_2$Ge$_2$Te$_6$ to form a coherent combination of 0 and $\pi$ JJ segments. The spin sensitivity of an Ising Josephson junction together with atomically thin magnetic tunneling barriers provides a route to the fabrication of novel superconducting and spintronic devices.

**Acknowledgments** We thank K. F. Mak, J. Shen, and Y. Otani for a fruitful discussion. The major part of the experiment was supported by ARO (W911NF-17-1-0574). The sample fabrication was supported by DOE QPress (DE-SC0019300). P.K. acknowledges support from the DoD Vannevar Bush Faculty Fellowship N00014-18-1-2877. H. I. acknowledges JSPS Overseas Research Fellowship and the Nakajima Foundation for support. L.T. N. and R. J. C. acknowledge the US Department of Energy, Division of Basic Energy Sciences, grant DE-FG02 98ER45706 for supporting the growth of the NbSe$_2$ crystals. Work done at Ames Lab (P. C. C. and N. H. J.) was supported by the U.S. Department of Energy, Office of Basic Energy Science, Division of Materials Sciences and Engineering. Ames Laboratory is operated for the U.S. Department of Energy by Iowa State University under Contract No. DE-AC02-07CH11358. N. H. J. was supported by the Gordon and Betty Moore Foundation's EPiQS Initiative through Grant GBMF4411.

**Author Contributions** H.I. fabricated the sample and analyzed the data. H.I. and K.F.H. performed the measurements. K.H. and D.S. performed TEM experiments. H.I. and P.K. conceived the experiment. F.P. developed the theoretical description. Y.J.S. provided the polymer and optimized transfer process. N.H.J. and P.C.C. grew and characterized single crystals of $Cr_2Ge_2Te_6$. L.T.N. and R.J.C. grew the $NbSe_2$ crystals. H.I. and P.K. wrote the manuscript with input from all other authors.

**References**

[1] Ryazanov, V. V. *et al*. Coupling of Two Superconductors through a Ferromagnet: Evidence for a π Junction. *Phys. Rev. Lett.* **86**, 2427-2430 (2001).

[2] Kontos T. *et al*. Josephson Junction through a Thin Ferromagnetic Layer: Negative Coupling. *Phys. Rev. Lett.* **89**, 137007 (2002).

[3] Blum, Y., Tsukernik, A., Karpovski, M. & Palevski, A. Oscillations of the Superconducting Critical Current in Nb-Cu-Ni-Cu-Nb Junctions. *Phys. Rev. Lett.* **89**, 187004 (2002).

[4] Oboznov, V. A., Bol'ginov, V. V., Feofanov, A. K., Ryazanov, V. V. & Buzdin, A. I. Thickness Dependence of the Josephson Ground States of Superconductor-Ferromagnet-Superconductor Junctions. *Phys. Rev. Lett.* **96**, 197003 (2006).

[5] Xi, X. *et al*. Ising pairing in superconducting $NbSe_2$ atomic layers. *Nat. Phys.* **12**, 139-143 (2016).

[6] Strambini, E. *et al*. A Josephson phase battery. *Nat. Nanotech.* **15**, 656-660 (2020).

[7] Goldobin, E. *et al*. D. Memory cell based on a φ Josephson junction. *Appl. Phys. Lett.* **102**, 242602 (2013).

[8] Gingrich, E. C. *et al*. Controllable 0–π Josephson junctions containing a ferromagnetic spin valve. *Nat. Phys.* **12**, 564-567 (2016).

[9] Menditto, R. *et al*. Tunable φ-Josephson junction ratchet. *Phys. Rev. E* **94**, 042202 (2016).

[10] Park, T., Ishizuka, H. & Nagaosa, N. Nonreciprocal transport of a super-Ohmic quantum ratchet. *Phys. Rev. B* **100**, 224301 (2019).

[11] Buzdin, A., Bulaevskii, L. & Panyukov, S. Critical-current oscillations as a function of the exchange field and thickness of the ferromagnetic metal (F) in an SFS Josephson junction. *JETP Lett.* **35**, 178-180 (1982).

[12] Buzdin, A. I. Proximity effects in superconductor-ferromagnet heterostructures. *Rev. Mod. Phys.* **77**, 935-976 (2005).

[13] Goldobin, E., Koelle, D., Kleiner, R. & Mints, R. G. Josephson Junction with a Magnetic-Field Tunable Ground State. *Phys. Rev. Lett.* **107**, 227001 (2011).

[14] Sickinger H. *et al*. Experimental Evidence of a ϕ Josephson Junction. *Phys. Rev. Lett.* **109**, 107002 (2012).

[15] Soloviev, I. I. *et al*. Beyond Moore's technologies: operation principles of a superconductor alternative. *Beilstein J. Nanotechnol.* **8**, 2689–2710, (2017).

[16] Kulagina, I. & Linder, J. Spin supercurrent, magnetization dynamics, and $\varphi$-state in spin-textured Josephson junctions. *Phys. Rev. B* **90**, 054504 (2014).

[17] Senapati, K., Blamire, M. G. & Barber, Z. H. Spin-filter Josephson junctions. *Nat. Mater.* **10**, 849-852 (2011).

[18] Pal, A., Barber, Z., Robinson, J. & Blamire, M. Pure second harmonic current-phase relation in spin-filter Josephson junctions. *Nat. Commun.* **5**, 3340 (2014).

[19] Massarotti, D. *et al*. Macroscopic quantum tunnelling in spin filter ferromagnetic Josephson junctions. *Nat. Commun.* **6**, 7376 (2015).

[20] Gong, C. *et al*. Discovery of intrinsic ferromagnetism in two-dimensional van der Waals crystals. *Nature* **546**, 265-269 (2017).

[21] Yabuki, N. *et al*. Supercurrent in van der Waals Josephson junction. *Nat. Commun.* **7**, 10616 (2016).

[22] Moodera, J. S., Hao, X., Gibson, G. A. & Meservey, R. Electron-Spin Polarization in Tunnel Junctions in Zero Applied Field with Ferromagnetic EuS Barriers. *Phys. Rev. Lett.* **61**, 637-640 (1988).

[23] Li, Y. F. *et al*. Electronic structure of ferromagnetic semiconductor $CrGeTe_3$ by angle-resolved photoemission spectroscopy. *Phys. Rev. B* **98**, 125127 (2018).

[24] Kim, M. *et al*. Strong Proximity Josephson Coupling in Vertically Stacked $NbSe_2$–Graphene–$NbSe_2$ van der Waals Junctions. *Nano Lett.* **17**, 6125-6130 (2017).

[25] Carteaux, V., Brunet, D., Ouvrard, G. & Andre, G. Crystallographic, magnetic and electronic structures of a new layered ferromagnetic compound $Cr_2Ge_2Te_6$. *J. Phys.: Condens. Matter* **7**, 69 (1995).

[26] Han, M.-G. *et al*. Topological Magnetic-Spin Textures in Two-Dimensional van der Waals $Cr_2Ge_2Te_6$. *Nano Lett*. **19**, 7859 (2019).

[27] Fletcher, J. *et al*. Penetration Depth Study of Superconducting Gap Structure of $2H$-$NbSe_2$. *Phys. Rev. Lett.* **98**, 057003 (2007).

[28] Bawden, L. *et al*. Spin–valley locking in the normal state of a transition-metal dichalcogenide superconductor. *Nat. Commun.* **7**, 11711 (2016).

[29] de la Barrera, S. C. *et al*. Tuning Ising superconductivity with layer and spin–orbit coupling in two-dimensional transition-metal dichalcogenides. *Nat. Commun.* **9**, 1427 (2018).

[30] Ménard, G. C. *et al*. Coherent long-range magnetic bound states in a superconductor. *Nat. Phys.* **11**, 1013-1016 (2015).

[31] Fulton, T. A. & Dunkleberger, L. N. Lifetime of the zero-voltage state in Josephson tunnel junctions. *Phys. Rev. B* **9**, 4760-4768 (1974).

**Methods**

**Crystal synthesis.** NbSe$_2$ crystals were grown from pre-cleaned elemental starting materials in an evacuated quartz glass tube, in a 700 to 650 °C temperature gradient, by iodine vapor transport. Cr$_2$Ge$_2$Te$_6$ was grown out of a ternary melt that was rich in the Ge-Te eutectic. High purity elements we placed into fritted alumina crucibles [32] in a ratio of Cr$_5$Ge$_{17}$Te$_{78}$, sealed in an amorphous silica ampoule under roughly 1/4 atmosphere of high purity Ar. The ampoule was heated over 5 hours to 900 °C, held at 900 °C for an additional 5 hours, and then cooled to 500 °C over 99 hours. At 500 °C, the excess liquid was separated from the Cr$_2$Ge$_2$Te$_6$ crystals with the aid of a centrifuge [32]. The single crystals grew as plates with basal plane dimensions of up to a cm and had mirrored surfaces perpendicular to the hexagonal c-axis (inset to Fig. S2). Low field magnetization on a bulk sample is shown in Fig. S2 and is consistent with a ferromagnetic transition near 65 K.

**Device Fabrication.** NbSe$_2$ and Cr$_2$Ge$_2$Te$_6$ crystals of the desired thickness were mechanically exfoliated onto a *p*-doped silicon chip terminated with 285 nm SiO$_2$. Exfoliated crystals were identified by optical contrast (some of them were separately characterized with atomic force microscope) in an argon-filled glove box. The NbSe$_2$/Cr$_2$Ge$_2$Te$_6$/NbSe$_2$ heterostructure was prepared by polymer-based dry transfer technique with the maximum process temperature of typically between 60 °C and 80 °C without taking out from the glove box to prevent degradation of the flakes [33]. The surface of the stack was examined by atomic force microscopy and/or scanning electron microscopy to identify the clean and atomically flat parts of the junction. Unnecessary parts were removed by reactive ion etching with fluorine gas using an electron (*e*)-beam (Elionix ELS-F125) patterned mask. For the SQUID device with field-calibration sensors, a double-layer resist was patterned by *e*-beam lithography followed by oblique deposition of aluminum. To form Al/Al$_2$O$_3$/Al junctions an *in-situ* oxidization process (1 mTorr, 10 min) was utilized between the *e*-beam evaporation of Aluminum. After the lift-off of the *e*-beam pattern, Ti/Au contacts were patterned by *e*-beam lithography using a polymethyl methacrylate (PMMA) mask followed by e-beam evaporation to contact both NbSe$_2$ and Al. Before this evaporation, the surface was *in-situ* cleaned by ion-milling.

**Preparation of the states for different switching current.** For the measurement in Fig. 3d, the switching current branch was prepared as follows: first, the device was measured by specific bias sweeps (from positive bias to negative bias, then returning to zero bias) at a magnetic field of 8.2 mT. Then we swept the magnetic field within the range between -0.4 mT and 0.4 mT to determine the switching current. The other switching current branch was characterized in the same manner, but the initial state was prepared by sweeping from positive bias to zero bias (at the same field of 8.2 mT).

**Lorentz transmission electron microscopy.** $Cr_2Ge_2Te_6$ flakes were obtained by mechanical exfoliation and transferred onto 50-nm-thick SiN membranes with holes supported by an Si frame in an Argon-filled glove box, by a similar procedure to the device fabrication. The samples were mounted on the liquid-Helium-cooling holder (ULTDT, Gatan) and inserted to the 300-kV field-emission TEM (HF-3300S, Hitachi High-Tech) specially designed for eliminating the magnetic field in the sample area. The Lorentz micrographs were taken using a defocusing condition in the electron optical system and the sample tilting condition (45°-30° tilt measured from the normal position for the optical axis) [34]. The thin flake (approximately 18-nm-thick) over the hole showed the disappearance of the stripe pattern near the holder temperature of 62 K, consistent with Curie temperature of $Cr_2Ge_2Te_6$.

**Theoretical model.**

The spectrum of monolayer $NbSe_2$ has Fermi pockets around the Γ point and around the K and K′ points, and the bands around the K and K′ points have sizable spin splitting $\Delta_{SOC} \simeq 100$ meV due to inversion-symmetry breaking spin-orbit interaction, which results in Ising superconductivity [29]. In contrast, bulk $NbSe_2$ has an inversion center between the layers, such that the SOC is opposite in even and odd layers destroying the spin-momentum locking. Because of weak interlayer hopping, however, bulk $NbSe_2$ can be thought of a stack of weakly coupled Ising superconductors with opposite spin polarization in even and odd layers. Tunneling in a Josephson junction predominantly originates from the layer adjacent to the junction and as a result the supercurrent will be carried by ICPs.

In our phenomenological model, we neglect any coupling between the layers. An effective low-energy Hamiltonian for the K and K′ valleys in a single layer can be written in Bogoliubov-de Gennes form in the Nambu-spinor basis $\psi = (\psi_\uparrow, \psi_\downarrow, \psi_\downarrow^\dagger, -\psi_\uparrow^\dagger)$ as

$$H_{SC} = v[p_\parallel - (p_F + p_{SOC}\sigma_z\lambda_z)]\tau_z + \Delta[\cos(\phi/2)\tau_x - \sin(\phi/2)\tau_y] \quad (1)$$

where $\sigma$, $\tau$ and $\lambda$ are Pauli matrices acting in spin, particle-hole and valley space, respectively. The spin-orbit coupling is opposite in the two valleys thus preserving time-reversal symmetry. The superconducting phase $\phi/2$ has opposite signs in the two leads, such that $\phi$ is the phase difference across the junction. For concreteness, we here assume that the SOC has the same sign both sides of the junctions. In the case of opposite signs, a similar argument for a $\phi$ junction can be made. We moreover assume the pairing strength $\Delta$ to be small compared to the spin splitting $2v\, p_{SOC}$ and hence the Cooper pairs consist of two electrons with opposite spins aligned with the $z$ direction. The magnetic layer is approximated by a single insulating band for each spin, whose energy bands are flat in two-dimensional momentum space. In the Nambu spinor basis $(d_\uparrow, d_\downarrow, d_\downarrow^\dagger, -d_\uparrow^\dagger)$, the Hamiltonian reads

$$H_{MI} = V\tau_z + J\vec{\sigma} \cdot \mathbf{n}, \quad (2)$$

where $V$ and $J$ denote the potential and exchange energy and $\mathbf{n}$ is a unit vector describing the direction of the magnetization. An extension to more complicated band structures is possible but will not qualitatively change our conclusions. The superconductors and the magnet are coupled by the hopping Hamiltonian

$$H_T = \sum_\sigma (t\psi_{L,\sigma}^\dagger d_\sigma + t\psi_{R,\sigma}^\dagger d_\sigma + h.c.) \quad (3)$$

where $t$ is positive. We now calculate the spectrum of Andreev bound states in the junction. For off-resonant tunneling, $t \ll V, J$, we can obtain an effective hopping between the left and right superconductor from second-order perturbation theory

$$H_{T,\text{eff}} = \tilde{t}\psi_{L,\sigma}^\dagger \psi_{R,\sigma} + h.c., \quad (4)$$

where the effective hopping strength is

$$\tilde{t} = \frac{t^2 V}{V^2 - J^2} - \frac{t^2 J \vec{\sigma} \cdot \mathbf{n}}{V^2 - J^2}. \quad (5)$$

If the junction is nonmagnetic, $J = 0$, we obtain $\tilde{t} = t^2/V$ and the Andreev spectrum simply is that of a narrow Josephson junction in a BCS superconductor.

$$E = \pm\Delta\sqrt{1 - D\sin^2\phi/2}, \quad (6)$$

where the transparency is $D = \pi^2 \tilde{t}^2 \nu^2/(1 + \pi^2 \tilde{t}^2 \nu^2)$ with the $\nu$ normal density of states in the superconductors. This is a regular Josephson junction whose ground state is at $\phi = 0$. For a purely magnetic junction with $\mathbf{n}$ along the $z$ axis, we instead obtain an effective hopping parameter

$$\tilde{t} = \frac{t^2 \sigma_z}{J}. \quad (7)$$

The hopping has a different sign for the two spin components and, hence, a Cooper pair tunneling across the junction acquires an additional minus sign with respect to a nonmagnetic junction. We can show this explicitly by doing a gauge transformation $\psi_{L,\downarrow} \to -\psi_{L,\downarrow}$ while leaving all other fermions invariant. In this new gauge the hopping is nonmagnetic $\tilde{t} \to t^2/J$ and the pairing term in the left superconductor changes sign $\Delta_L = \langle\psi_{L,\sigma}^\dagger\psi_{L,\sigma}^\dagger\rangle \to -\Delta_L$ while the remaining terms are unchanged. This shows that we obtain the same spectrum as before but with a $\pi$ phase shift

$$E = \pm\Delta\sqrt{1 - D\sin^2(\phi + \pi)/2}. \quad (8)$$

Hence the ground state of the Josephson junction is at $\phi = \pi$. In fact, the system always forms a $\pi$ junction when $J > V$ as was first noted in ref. 35.

Now we consider a junction with magnetization along the $x$ direction so that scattering in the junction can result in spin flips. Due to the strong spin-orbit coupling, however, the band structure in the superconductor is helical, meaning that at any particular in-plane momentum there is only one spin component at the Fermi level. Thus, if a spin flip occurs in the barrier the other spin component has a large momentum mismatch when entering the superconductor. The latter therefore acts as a hard wall for flipped spins as long as the superconductor-magnet interface is sufficiently clean, such that scattering approximately conserves the in-plane momentum. Andreev reflection can therefore only happen after an even number of spin flips in the barrier, which means the supercurrent is an even function of $J$ in this case and all spin dependence drops out. This implies in particular that hopping has the same sign for electrons with different spins and hence the junction always has a ground state at zero.

Now let us assume that the magnet is inhomogeneous and there are regions with magnetization along $z$ and $x$. This means critical current changes sign as a function of the in-plane position. When the length scale of the spatial variations is smaller than the Josephson screening length the critical current is simply the spatial average of the current. As a simple model, we assume a fraction $\lambda$ of the plane favors a $\pi$ junction described by Eq. (8). The remaining fraction (1-$\lambda$) is instead described by Eq. (6). Note that the latter also includes a conventional Josephson current due to electron near the $\Gamma$ point. In Fig. 4d we plot the spectrum of the Josephson junction when the transparency is $D$=0.75 in both regions and $\lambda$ =0.53. See the SI for the microscopic description of the theoretical model.

**References**

[32] Canfield, P. C., Kong, T., Kaluarachchi, U. S. & Jo, N. H. Use of frit-disc crucibles for routine and exploratory solution growth of single crystalline samples. *Philos. Mag.* **96**, 84-92 (2016).

[33] Son, S. *et al*. Strongly adhesive dry transfer technique for van der Waals heterostructure. *2D Mater.* **7**, 041005 (2020).

[34] Harada, K. Lorentz microscopy observation of vortices in high-$T_C$ superconductors using a 1-MV field emission transmission electron microscope. *Microscopy* **62**, S3-S15 (2013).

[35] Bulaevskii, L.N., Kuzii, V.V. & Sobyanin, A.A. On possibility of the spontaneous magnetic flux in a Josephson junction containing magnetic impurities. *Solid State Commun.* **25**, 1053-1057 (1978).

# Figures

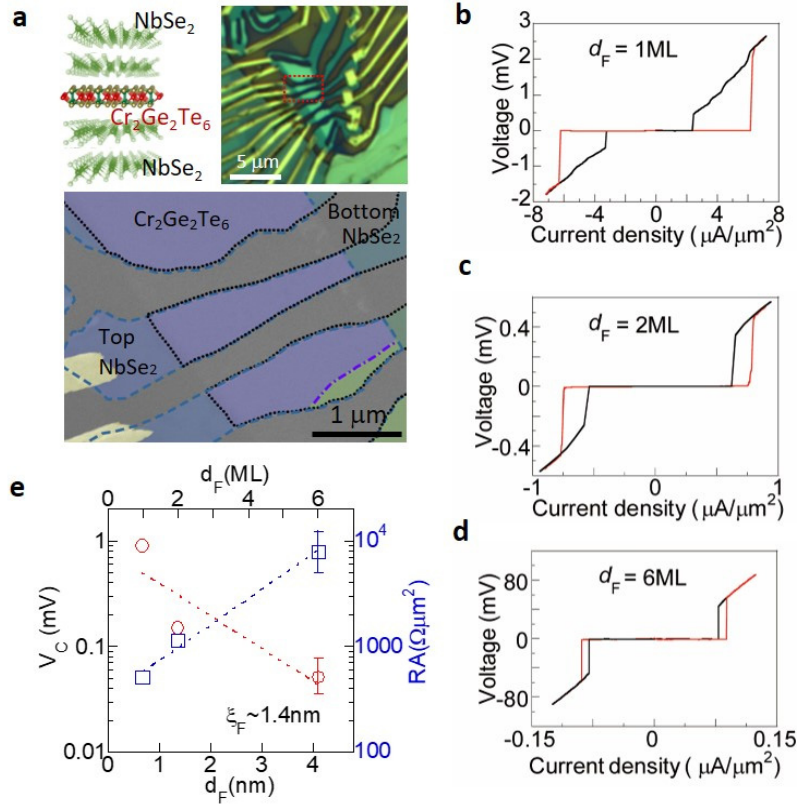

**Fig. 1. Josephson coupling of the NbSe₂/Cr₂Ge₂Te₆/NbSe₂ van der Waals junction. a,** Schematic (top left panel), optical micrograph (top right panel), and false color scanning electron microscope (SEM) image (bottom panel) of NbSe$_2$/Cr$_2$Ge$_2$Te$_6$/NbSe$_2$ Josephson junctions (JJs). In the schematic, top and bottom pairs of layers represent NbSe$_2$ and the middle layer represents Cr$_2$Ge$_2$Te$_6$. In the optical micrograph, the gold contact leads appear as yellow lines and the JJ appears in green. The red dotted line indicates the area of the SEM image. In the SEM image, the areas of the top and bottom NbSe$_2$ flakes are indicated by the blue dashed line and the black dotted line, respectively. The Cr$_2$Ge$_2$Te$_6$ and NbSe$_2$ layers are indicated by purple and green. The dash-dot line indicates the boundary of Cr$_2$Ge$_2$Te$_6$. **b-d,** $V$-$J$ characteristics of the JJs for the Cr$_2$Ge$_2$Te$_6$ barrier thicknesses of 1 ML, 2 ML, and 6 ML. The junction area $A$ is ~ 0.9 μm$^2$, ~ 4 μm$^2$, and ~ 1.4 μm$^2$ respectively. The red lines depict the transition from the superconducting state to the normal state whereas the black ones depict the transition from the normal state to the superconducting state. **e,** The characteristic voltage (switching current normal resistance product) and resistance-area product of the junction with different barrier thickness. The thickness of the barrier is shown in units of both nanometers and the number of layers (ML).

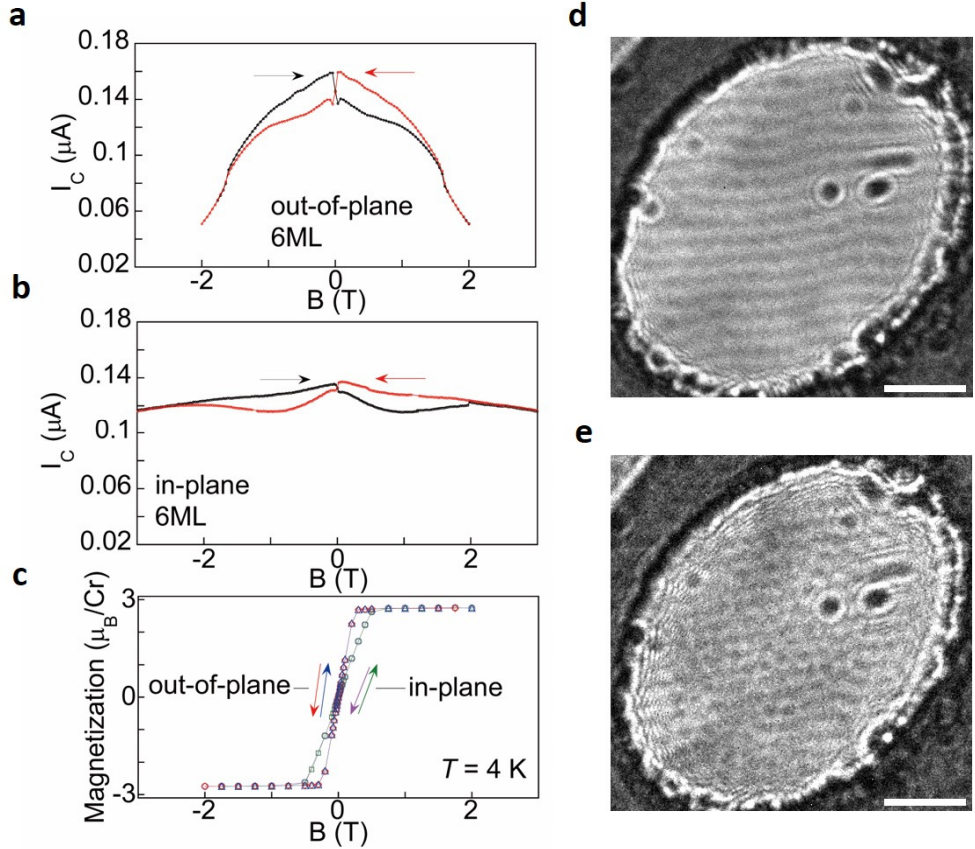

**Fig.2. The magnetic characteristics of the Josephson Junction and Cr$_2$Ge$_2$Te$_6$. a, b,** The switching current of the NbSe$_2$/Cr$_2$Ge$_2$Te$_6$/NbSe$_2$ junction with a 6 ML Cr$_2$Ge$_2$Te$_6$ barrier at the temperature of 0.3 K as a function of the applied magnetic field (**a**) perpendicular (**b**) and parallel to the 2D layer. The red lines indicate the magnetic field scan from positive polarity to negative polarity and the black ones from negative polarity to positive polarity. **c,** The bulk magnetization of a Cr$_2$Ge$_2$Te$_6$ crystal measured by a SQUID magnetometer at 4 K. The red diamonds and blue triangles indicate the data with perpendicular field, and the purple circles and green triangles indicate the data with in-plane field. The arrows indicate the sweep direction for **a-c**. **d,** The magnetic domain structure of a Cr$_2$Ge$_2$Te$_6$ flake measured by Lorentz microscopy without an applied magnetic field. The thickness of the flake is 18 nm. The sample was tilted by 45 degrees from the incident electron beam direction to image an out-of-plane magnetization pattern at the temperature of 25 K. The black and white contrast in the central oval-shape area indicates magnetic domain wall contrast caused by the opposite magnetization. The oval-shape is the area where Cr$_2$Ge$_2$Te$_6$ is suspended over a hole in the SiN membrane support. **e,** Lorentz micrograph of the same sample with **d** but taken with a different thermal cycle. Unlike the simple stripe domains shown in **d**, bubble-like domains appear to break up the stripe-like domains. The scale bars correspond to 0.5 μm.

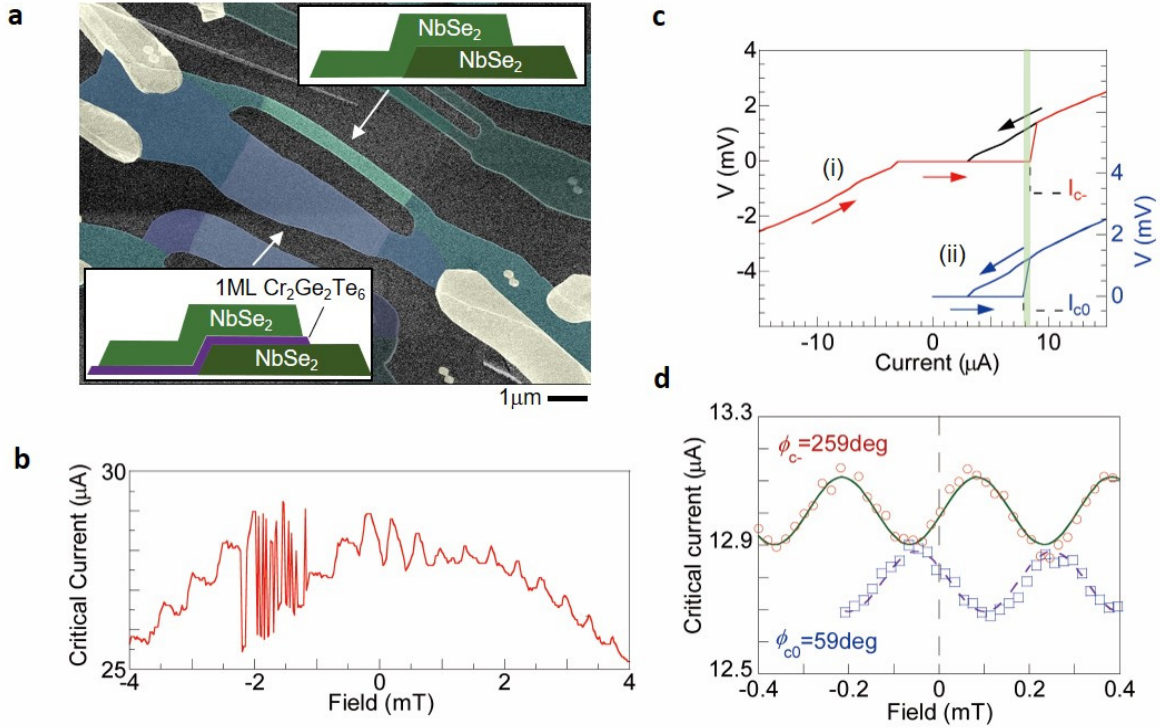

**Fig. 3. Josephson phase measurement of the vdW SQUID devices with ultrathin magnetic insulator junction. a,** False-color SEM image of the NbSe$_2$ SQUID device with and without a 1 ML Cr$_2$Ge$_2$Te$_6$ barrier (Cr$_2$Ge$_2$Te$_6$-SQUID). The inset shows a schematic of cross-section of the junctions indicated by arrows. The bare NbSe$_2$/NbSe$_2$ interface can still form an intrinsic JJ due to the residual interfacial strain [21]. **b,** The SQUID oscillates with telegram-like critical current between the fields of -1.1 mT and -2.2 mT, which implies the presence of two metastable state. The SQUID loop area (≈ 4.5 μm$^2$) is the sum of the empty area of the SQUID ≈ 2.5 μm$^2$ and the area for screening ≈ 2 μm$^2$ agrees with the area calculated from the oscillation period ~ 4-6 μm$^2$. **c,** The sweep-dependent critical currents for Cr$_2$Ge$_2$Te$_6$-SQUID measured at T = 1.1 K. The red line labeled "(i)" depicts the sweep from negative bias (current) to positive bias, with switching current $I_{c-}$. The black line depicts the sweep from positive bias to zero bias. The blue line labeled "(ii)" depicts the sweep from positive bias to zero bias followed by a sweep from zero bias to positive bias, whose switching current $I_{c0}$ is lower than $I_{c-}$. The green area indicates the difference of the switching currents. **d,** The Cr$_2$Ge$_2$Te$_6$-SQUID oscillation. The Cr$_2$Ge$_2$Te$_6$-SQUID loop area (≈ 7 μm$^2$) is the sum of the empty area of the SQUID ≈ 3 μm$^2$ and the area for screening ≈ 4.5 μm$^2$ agrees with the area calculated from the oscillation period ~ 7 μm$^2$. The field is calibrated by three different sizes of aluminum SQUIDs located 0.1 mm away from the Cr$_2$Ge$_2$Te$_6$-SQUID. This allows for precise calibration of zero field and thus measurement of the phase of the SQUID oscillation. Red circles and blue rectangles represent $I_{c-}$ and $I_{c0}$ measured at T = 0.9 K. Each branch has a different phase, indicated by a fit to $I_{c-}$ and $I_{c0}$ with the sine curves shown in green and purple.

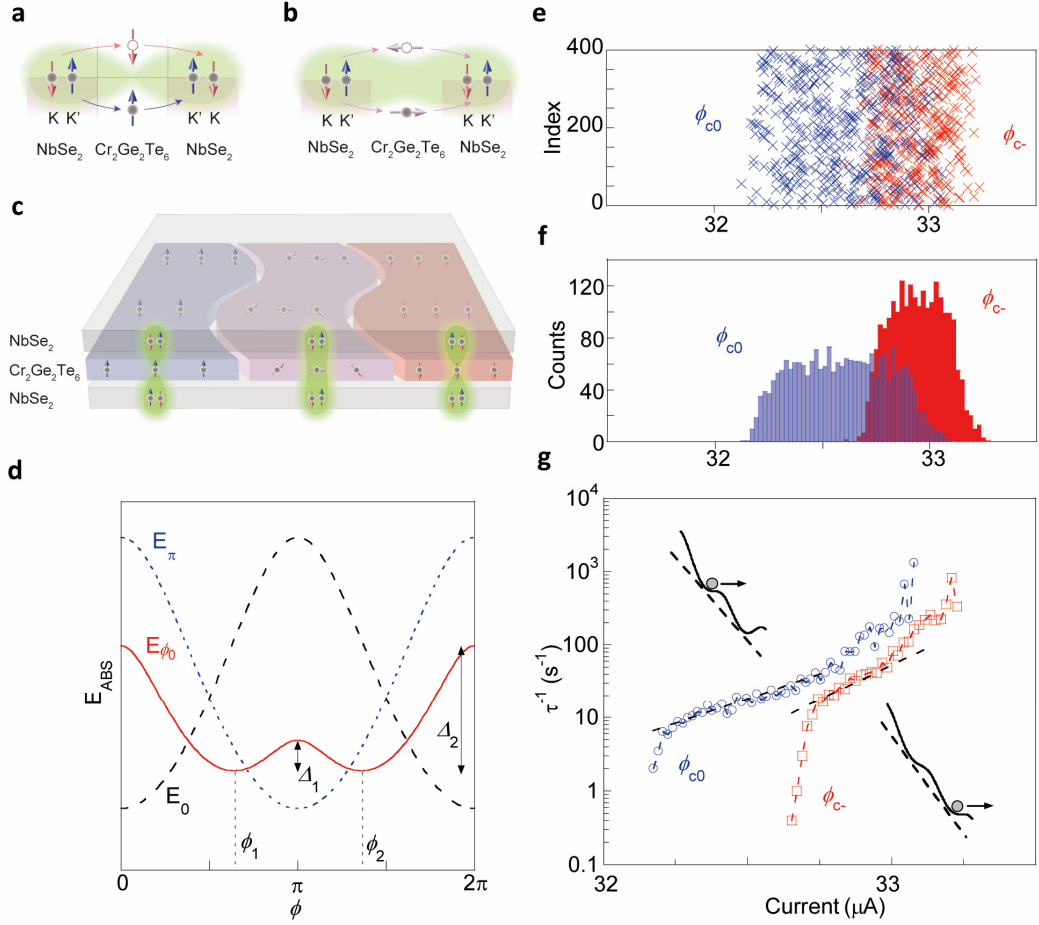

**Fig. 4. Ising-Cooper-pair coupling through the all-crystalline vdW magnetic Josephson junction. a-c,** Illustration of the Ising-Cooper-pair coupling through the magnetic tunneling junction for perpendicular magnetization (**a**), for in-plane magnetization (**b**), and for magnetic structure (**c**). For perpendicular magnetization, Ising Cooper pairs can tunnel via the spin-dependent energy levels without spin flip, forming a $\pi$ phase junction. For in-plane magnetization, tunneling occurs with a spin flip, which requires zero phase. With the magnetic domain structure of $Cr_2Ge_2Te_6$, the Josephson junction consists of segments of $\pi$ phase and 0 phase. **d,** Washboard potential for the $\phi$ junction (indicated by solid red line), zero junction (dashed black line), and $\pi$ junction (dotted blue line). Nontrivial Josephson phase $\phi_1$ and $\phi_2$ appear as the minimum energy for $\phi$ state (the red line is calculated using $\lambda = 0.53$ and $D = 0.75$). **e-g,** switching current distribution for the $\phi$ state at $T = 1.1$ K. Different switching current is repeatedly acquired with different sweeping, as depicted in Fig. 2c. The lower and higher critical currents are depicted in blue and red. **e,** Switching currents with different sweeping events **f,** Histogram of switching currents in 0.02 μA bins. **g,** Escape rate $\tau^{-1}$ as a function of the bias current. The lifetime $\tau$ has a different slope reflected by the different metastable states of the tilted washboard potential shown in **d**, and the potential barrier for each escaping event is schematically illustrated in the insets.

*Supplementary information*

*for*

**Van der Waals Heterostructure Magnetic Josephson Junction**

H. Idzuchi[1], F. Pientka[1,2], K.-F. Huang[1], K. Harada[3], Ö. Gül[1], Y. J. Shin[1], L. T. Nguyen[4], N. H. Jo[5,6], D. Shindo[3], R. J. Cava[4], P. C. Canfield[5,6], and P. Kim[1*]

[1] *Department of Physics, Harvard University, Cambridge, MA 02138, USA*

[2] *Institut für Theoretische Physik, Goethe-Universität, 60438 Frankfurt am Main, Germany*

[3] *Center for Emergent Matter Science (CEMS), RIKEN, Wako, Saitama 351-0198, Japan*

[4] *Department of Chemistry, Princeton University, Princeton, NJ 08540, USA*

[5] *Department of Physics and Astronomy, Iowa State University, Ames, IA 50011, USA*

[6] *Ames Laboratory, Iowa State University, Ames, IA 50011, USA*

**Supplementary Note S1. Magnetic-field hysteresis of critical current in the junction with thick $Cr_2Ge_2Te_6$ barrier.**

We have measured the critical current of $NbSe_2/Cr_2Ge_2Te_6/NbSe_2$ junction with a 6 ML $Cr_2Ge_2Te_6$ barrier as a function of the magnetic field in perpendicular to the 2D layer. The data were taken both for decreasing (the red lines and arrows) and increasing (the black lines and arrows) magnetic fields, as shown in Figure 2a-b. The magnetic-field hysteresis of critical current shows a clear hysteresis. The characteristic field strength of the hysteresis in critical current differs from the hysteresis of magnetization in $Cr_2Ge_2Te_6$ alone, as the magnetization curves measured by SQUID is shown in Figure 2c. To study the origin of the hysteresis, we have measured the magnetic structure of $Cr_2Ge_2Te_6$ by means of a transmission electron microscope with Lorentz mode (Lorentz transmission electron microscopy, see Methods for the detail). We have observed stripe magnetic structures (Fig.2d) and bubble-like one (Fig.2e) in the Lorentz micrographs. We did not see the pattern when the sample is placed perpendicular to the electron beam, and the pattern appears when the sample tilted, indicating that the magnetic structure exhibits out-of-plane magnetic moment. The stripe pattern appears dominantly, and bubble-like one appeared in the same sample with different thermal cycle, implying the later one is in metastable state.

Figure S1a shows the possible magnetic configuration in NbSe$_2$/Cr$_2$Ge$_2$Te$_6$(6ML)/NbSe$_2$ junction under the perpendicular magnetic field. When the field is large, dense vortices in NbSe$_2$ can form a uniform magnetic structure in Cr$_2$Ge$_2$Te$_6$. With decreasing the strength of the field, the density of vortex decreases. The vortex array makes the energy cost to form the stripe pattern larger, and the magnetic structure with the hexagonal-like pattern becomes more stable (Fig.S1c). With a small external magnetic field, the magnetic field is shielded by NbSe$_2$ (Meissner effect), and the magnetic domain of Cr$_2$Ge$_2$Te$_6$ should prefer the stripe pattern (Fig.S1d). With increasing the field, the NbSe$_2$ forms vortices and the energy to form bubble-like pattern decreases while there is an energy barrier for a switching between the two types of the structures. With increasing the field, it finally leads to the uniform magnetization state again (Fig.S1b).

The magnetization of bulk Cr$_2$Ge$_2$Te$_6$ does not show notable hysteresis (Fig.2c) in consistent with earlier reports [20,25]. This indicates small magnetic structure, which is averaged out in the magnetization measurement of a crystal, may play a role in switching current of Josephson junction.

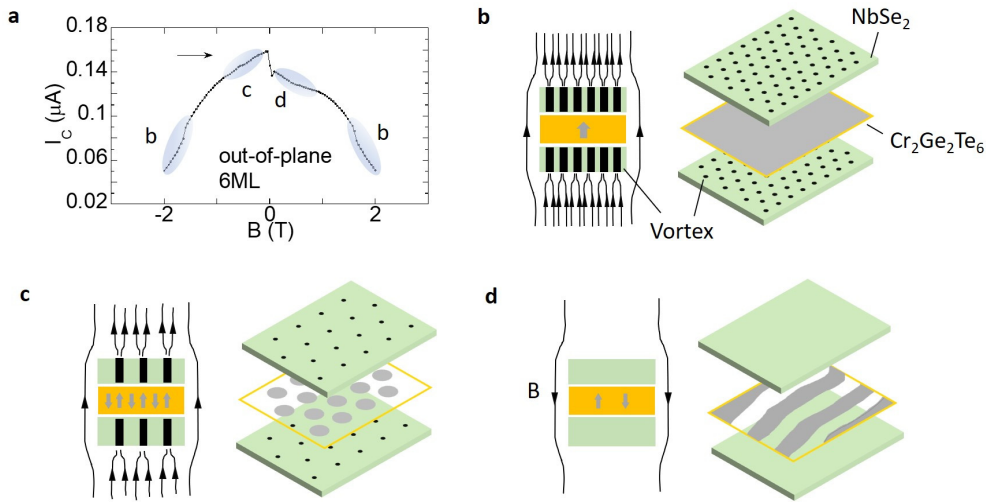

**Figure S1| Possible configuration of magnetic domain structure in Cr$_2$Ge$_2$Te$_6$ with vortices in NbSe$_2$. a,** The critical current of NbSe$_2$/Cr$_2$Ge$_2$Te$_6$/NbSe$_2$ junction with a 6 ML Cr$_2$Ge$_2$Te$_6$ barrier at the temperature of 0.3 K as a function of the magnetic field in perpendicular to 2D layer. The arrows indicate sweep direction. The labels **"b"** to **"d"** attached to the filled areas indicates the field regions corresponding to the configurations of (**b**) – (**d**). **b-d,** Preferential magnetic structure in Cr$_2$Ge$_2$Te$_6$ with vortex pattern in NbSe$_2$. Green and orange boxes indicate NbSe$_2$ and Cr$_2$Ge$_2$Te$_6$, respectively. Black closed area and dot indicate superconducting vortex. The black lines attached to triangles indicate magnetic field lines. The gray arrows in the red box indicate magnetization direction of Cr$_2$Ge$_2$Te$_6$. Gray closed area and dot in right panels indicate magnetic domain for up magnetization. As the vortex can penetrate and focus the magnetic field near the core, the part of the Cr$_2$Ge$_2$Te$_6$ exposed to the strong field should be magnetized along that direction. (**d**) is depicted for a Meisner phase.

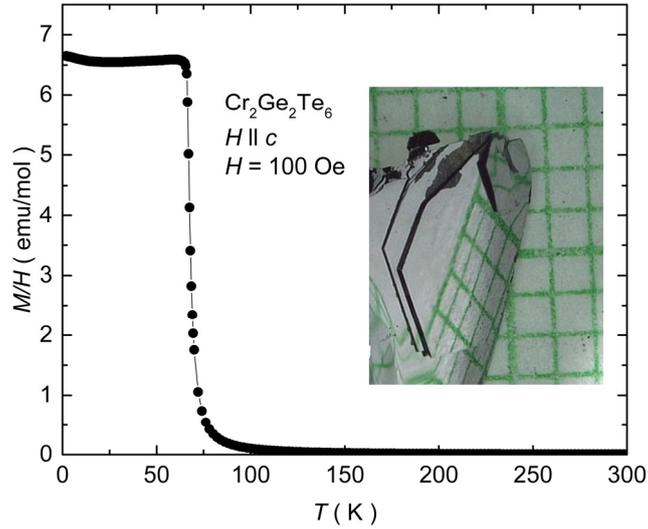

**Figure S2| Cr$_2$Ge$_2$Te$_6$ single crystal.** Temperature variation of low field magnetization. Inset shows photograph of a faceted single crystal on mm-grid paper. The mirrored face of the crystal is reflecting the mm-grid in the image.

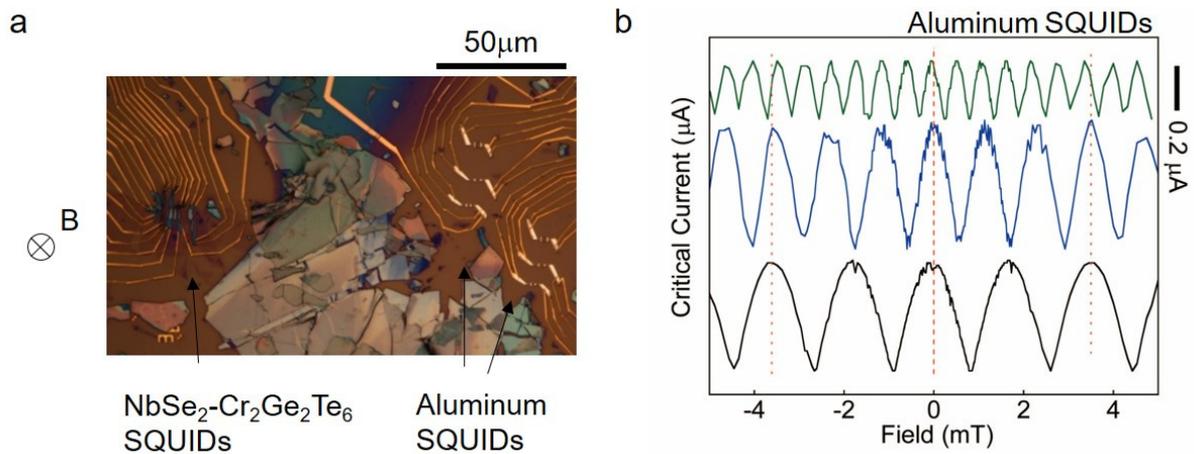

**Figure S3| Aluminum SQUIDs for magnetic field calibration. a,** Optical micrograph of the Cr$_2$Ge$_2$Te$_6$-SQUID with aluminum SQUIDs for calibration of magnetic field. **b,** Critical-current oscillation for the three aluminum SQUIDs with different sensor areas, measured at $T = 0.3$ K. The plots are offset in the vertical direction for clarity.

**Supplementary Note S2. Model Calculation**

In order to corroborate our theoretical arguments in the main paper, we now evaluate the energy-phase relation of a magnetic-insulator Josephson junction based on a concrete microscopic model. We assume translational invariance in the plane perpendicular to the current, which flows along the $z$ direction. Hence our model becomes effectively one-dimensional and the two in-plane momentum components $\mathbf{p}_\parallel$ enter as parameters of the Hamiltonian. NbSe$_2$ has an inversion center between two layers that form the unit cell. As a consequence, each layer has strong spin-orbit coupling, which has opposite sign in even and odd layers. This leads to a large spin-orbit splitting $\Delta_{SOC}$ of the Fermi pockets around the the $K$ and $K'$ points within each layer, i.e., a locking of spin and layer degrees of freedom with spins quantized along the $z$ direction [5]. Because of the weak interlayer hopping in NbSe$_2$, we can consider the electrons from even and odd layers as forming approximately independent bands. The Josephson effect is dominated by contributions from the first layer on each side of the magnetic barrier so we can focus only on the odd-layer band around the $K$ and $K'$ points. Any coupling from the other spin band at the $K$ and $K'$ points as well as from the pocket around the $\Gamma$ point, which does not exhibit spin-layer locking, will yield an additional contribution to Josephson energy with energy minimum at $\phi=0$, but will not change our general conclusions.

The strong spin-orbit coupling in the odd layers separates the two spins in momentum space around the $K$ and $K'$ points. Because the spin splitting exceeds the out-of-plane hopping strength, the bandwidth along the out-of-plane momentum $p_z$ is smaller than the energy separation between different spin bands and we therefore expect the odd-layer Fermi surface sheets for spin up and down to occur at different in-plane momenta. This assumption is corroborated by DFT calculations [28]. Moreover, the spin splitting also exceeds pairing strength, $\Delta_{SOC} \gg \Delta$, which means the Cooper pairs have spins quantized along the $z$ axis. As a consequence, we can consider the Hamiltonian $H(\mathbf{p}_\parallel)$ at a momentum $\mathbf{p}_\parallel$ on the Fermi surface as effectively spinless and ignore scattering to the spin which does not cross the Fermi level at that particular in-plane momentum.

**i) Equal spins on both sides of the junction.**

We assume for now an orientation of the NbSe$_2$ such that the Fermi surfaces with the same spin polarization on both sides of the junction are aligned in momentum space. We will later comment on the opposite case. The four Fermi surface sheets can labeled by $\rho=\pm 1$ which equals $+1$ on the inner (outer) Fermi surface of the $K$ ($K'$) valley and $-1$ otherwise. The corresponding Hamiltonian in the continuum approximation is given by

$$H(p_\parallel) = \left[\frac{p_z^2}{2m_z} + \frac{p_\parallel^2}{2m_\parallel} - \mu - \rho\Delta_{SO}\sigma_z)\right]\tau_z$$
$$+ \delta(z)(V\tau_z + J\vec{\sigma}\cdot\mathbf{n}) + \Delta\cos[\phi(z)]\tau_x - \Delta\sin[\phi(z)]\tau_y,$$

(S1)

where $\mu$ is the chemical potential and the superconducting phase $\phi(z)=\text{sign}(z)\phi/2$ jumps at $z=0$, the location of the barrier. The narrow magnetic barrier is modeled as a $\delta$-function and includes potential scattering with strength $V$ and magnetic scattering with strength $J$, which have both dimensions of velocity. The Hamiltonian has translational invariance in the plane and $p_{||} = (p_x^2 + p_y^2)^{1/2}$ is therefore conserved. The spin-orbit splitting is $\Delta_{SOC} > |p_{||}^2/2m_{||} - \mu|$ such that system behaves as a superconductor for one spin and as an insulator for the other spin.

The scattering matrix of the barrier for states at the Fermi energy is given by

$$S_B = \begin{pmatrix} r & t \\ t & r \end{pmatrix} = \frac{1}{1-i\beta}\begin{pmatrix} i\beta & 1 \\ 1 & i\beta \end{pmatrix}, \tag{S2}$$

$$\beta = -\frac{(V\tau_z + J\vec{\sigma}\cdot\mathbf{n})}{v_z}, \tag{S3}$$

where $v_z$ is the Fermi velocity in the $z$ direction. At subgap energies, the superconducting leads have zero transmission and the corresponding scattering matrix is given by

$$S_{SC} = \begin{pmatrix} r_L & 0 \\ 0 & r_R \end{pmatrix}, \tag{S4}$$

where $r_{L/R}$ are the reflection matrix elements of the left and right lead, which depend on the spin of the incoming electrons. The reflection amplitudes can be approximated by

$$r_{L/R} = \frac{(1+\rho\sigma_z)}{2}\tau_x e^{i\alpha \pm i\alpha\phi/2\tau_z} - e^{i\gamma}\frac{(1-\rho\sigma_z)}{2}, \tag{S5}$$

where the first term accounts for Andreev reflection of spin $\sigma_z = \rho$ with $\cos\alpha = E/\Delta$ and the second term describes normal reflection of spin $\sigma_z = -\rho$ with $\gamma$ the reflection phase. We can obtain the energy spectrum of Andreev bound states in the junction from the condition

$$\det(1 - S_{SC}S_B) = 0. \tag{S6}$$

We now distinguish two cases depending on the magnetization direction in the magnetic insulator.

(i) *Out-of-plane magnetization.* For $\mathbf{n}=\hat{z}$ the Hamiltonian conserves spin and from Eq.(S6) we obtain in the $\sigma_z = \rho$ channel

$$\frac{V^2 - J^2}{v_F^2} + \cos\phi = \left(1 + \frac{V^2 - J^2}{v_F^2}\right)\cos(2\alpha) + \rho\frac{2J}{v_F}\sin(2\alpha), \tag{S7}$$

which yields four energy levels when both valleys are taken into account ($\rho=\pm 1$). Closed analytical expressions for the energies can be easily obtained but are not very illuminating. We instead plot the result in Fig. S4, which shows the energy spectrum as well as the ground state energy $E_{gs}^\perp(\phi)$ obtained by summing over the two negative energy solutions. For sufficiently large values of $J$, the Josephson junction has its ground state at $\phi = \pi$. For $V = 0$ the threshold for a $\pi$-junction is $J/v_F \gtrsim 0.8$.

(ii) *In-plane magnetization.* For $\mathbf{n}=\hat{x}$ the spins can flip at the barrier. However, in the limit of large $\Delta_{SO}$, the superconductors are essentially hard walls for electrons with spin $\sigma_z=-\rho$ and thus spin flip scattering is suppressed.

Retaining only spin conserving scattering we obtain

$$\frac{V^2}{v_F^2} + \cos\phi = \left(1 + \frac{V^2}{v_F^2}\right)\cos(2\alpha), \tag{S8}$$

which is identical to Eq. (S7) with $J=0$. One can explicitly verify this result by evaluating the full expression in Eq. (S6) and subsequently taking the limit $\gamma \to 0$, which corresponds to hard-wall reflection of spins $\sigma_z = -\rho$ in the limit $\Delta_{SOC} \to \infty$. The valley degenerate energies are given by

$$E^{\parallel}(\phi) = \pm\Delta\sqrt{\frac{1 + 2(V/v_F)^2 + \cos\phi}{2 + 2(V/v_F)^2}}, \tag{S9}$$

and the ground state energy $E_{gs}^{\parallel}(\phi) = -2|E^{\parallel}(\phi)|$ has a global minimum at $\phi=0$.

(iii) *Magnetization with arbitrary angle.* In the case of a magnetization with both in-plane and out-of-plane components, $\mathbf{n} = (n_x, 0, n_z)$, the in-plane part, which leads to spin flip scattering, can again be ignored. This case therefore reduces to Eq.(S7) with the replacement $J \to Jn_z$, i.e., it smoothly interpolates between the cases (i) and (ii). As the magnetization is tilted towards the $x$ - $y$ plane, $n_z$ becomes smaller effectively reducing the magnetic scattering amplitude. For sufficiently small out-of-plane magnetization the Josephson energy will always have a global minimum at $\phi=0$.

The total Josephson energy depends on the spatial distribution of magnetic domains. If we assume that a fraction $\lambda$ of the plane has an out-of-plane magnetization and the rest has an in-plane magnetization, the total ground state energy is

$$E_{gs}(\phi) = \lambda E_{gs}^{\perp}(\phi) + (1-\lambda) E_{gs}^{\parallel}(\phi). \tag{S10}$$

Here we assume for simplicity a magnetization that is either in plane or out of plane, although a more complicated magnetic texture can be easily included. The ground state phase different for $\lambda = 0.9$ is plotted in Fig.S5. There is clearly an extended region of parameter phase with a nonzero phase difference.

ii) **Opposite spins on both sides of the junction.**

In our model, we have so far assumed that the NbSe$_2$ Fermi surfaces have the same spin in the first layer on both sides of the junction. In the case when the spins are opposite, we can consider the same model with the spin orbit coupling replaced by $\Delta_{SOC} \to \Delta_{SOC}$ sgn($z$). Cooper pair tunneling across the barrier now requires a spin flip, i.e., the magnetization must be in-plane. A similar calculation as above, shows that the ground state energy has a minimum at $\phi=\pi$ in a sizable fraction of parameter space. Note that in this case we need to choose $\gamma \neq 0$, because in the presence of a hard wall (i.e., $\gamma=0$) the wavefunction would have a node at the $\delta$-function, which would render spin flips impossible and result in zero Josephson current.

When the magnetization is out of plane, our simple model yields no Josephson current because of spin conservation. In that case, other contributions, e.g., from the $\Gamma$ point or due to spin mixing around the $K$ and $K'$ point, would presumably lead to a ground state at zero phase. In conclusion, there is a similar competition between zero- and $\pi$-junctions that can conceivably result in an overall ground state at a nontrivial phase $\phi \neq 0, \pi$.

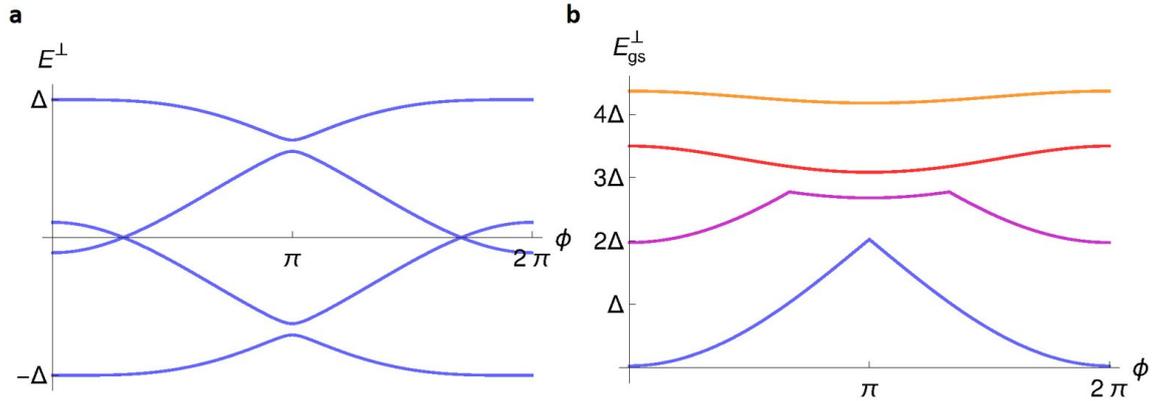

**Figure S4|** **a**, Energy spectrum of Andreev bound states for out-of-plane magnetization with $J/v_F=0.9$ and $V/v_F=0.1$. **b**, Ground state energy for $V=0$ and $J/v_F=0,0.5,1,2$ from bottom to top (curves are shifted vertically for clarity).

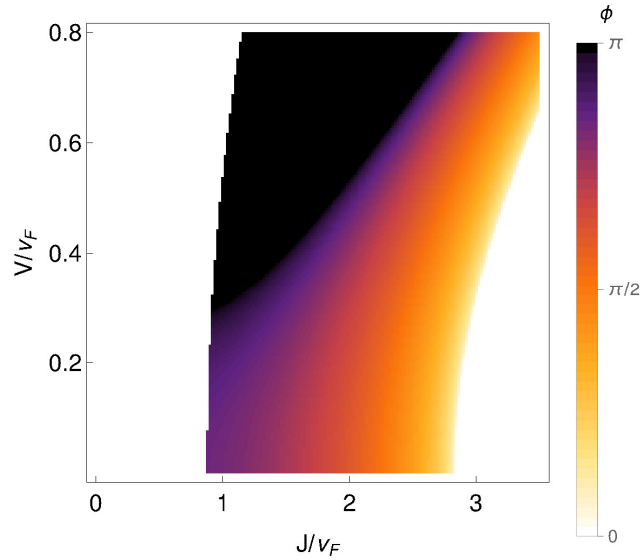

**Figure S5|** Phase difference $\phi$ of the ground state energy in Eq.(S10) for $\lambda=0.9$. The white and black regions correspond to phases zero and $\pi$. The colored areas indicate a nontrivial phase difference across the junction in the ground state.